\documentclass[twocolumn,english,aps,prl,showpacs,floatfix,superscriptaddress]{revtex4}
\usepackage[T1]{fontenc}
\usepackage[latin9]{inputenc}
\usepackage{amsmath}
\usepackage{amssymb}
\usepackage{graphicx}
\usepackage{float}

\makeatletter
\@ifundefined{textcolor}{}
{%
 \definecolor{BLACK}{gray}{0}
 \definecolor{WHITE}{gray}{1}
 \definecolor{RED}{rgb}{1,0,0}
 \definecolor{GREEN}{rgb}{0,1,0}
 \definecolor{BLUE}{rgb}{0,0,1}
 \definecolor{CYAN}{cmyk}{1,0,0,0}
 \definecolor{MAGENTA}{cmyk}{0,1,0,0}
 \definecolor{YELLOW}{cmyk}{0,0,1,0}
}

\newcommand*{\balancecolsandclearpage}{%
  \close@column@grid
  \clearpage
  \twocolumngrid
}

\@ifundefined{definecolor}
 {\usepackage{color}}{}
\usepackage{siunitx}

\makeatother

\usepackage{babel}
\begin{document}
\flushbottom

\title{Macroscopic two-dimensional polariton condensates}

\author{Dario~Ballarini}

\affiliation{CNR NANOTEC - Istituto di Nanotecnologia, Via Monteroni, 73100 Lecce, Italy}

\author{Davide~Caputo}

\affiliation{CNR NANOTEC - Istituto di Nanotecnologia, Via Monteroni, 73100 Lecce, Italy}
\affiliation{University of Salento, Via Arnesano, 73100 Lecce, Italy}

\author{Carlos S\'anchez Mu\~noz}

\affiliation{Departamento de F\'{i}sica Te\'orica de la Materia Condensada, Universidad Aut\'onoma de Madrid, 28049 Madrid, Spain}

\author{Milena~De~Giorgi}

\affiliation{CNR NANOTEC - Istituto di Nanotecnologia, Via Monteroni, 73100 Lecce, Italy}

\author{Lorenzo~Dominici}
\affiliation{CNR NANOTEC - Istituto di Nanotecnologia, Via Monteroni, 73100 Lecce, Italy}

\author{Marzena~H.~Szyma\'nska}
\affiliation{Department of Physics and Astronomy, University College London,
Gower Street, London WC1E 6BT, United Kingdom}

\author{Kenneth~West}
\affiliation{PRISM, Princeton Institute for the Science and Technology of Materials, Princeton Unviversity, Princeton, NJ 08540}

\author{Loren~N.~Pfeiffer}
\affiliation{PRISM, Princeton Institute for the Science and Technology of Materials, Princeton Unviversity, Princeton, NJ 08540}

\author{Giuseppe~Gigli}

\affiliation{CNR NANOTEC - Istituto di Nanotecnologia, Via Monteroni, 73100 Lecce, Italy}

\affiliation{University of Salento, Via Arnesano, 73100 Lecce, Italy}

\author{Fabrice~P.~Laussy}

\affiliation{University of Wolverhampton, Faculty of Science \& Engineering, Wulfruna St, Wolverhampton WV1 1LY, UK}

\affiliation{Russian Quantum Center, Novaya 100, 143025 Skolkovo, Moscow Region,
Russia}

\author{Daniele~Sanvitto}

\affiliation{CNR NANOTEC - Istituto di Nanotecnologia, Via Monteroni, 73100 Lecce, Italy}
\affiliation{INFN, Sez. Lecce, 73100 Lecce, Italy}

\begin{abstract}

We report a record-size, two-dimensional polariton condensate of a fraction of a millimeter radius free from the presence of an exciton reservoir. This macroscopically occupied state is formed by the ballistically expanding polariton flow that relaxes and condenses over a large area outside of the excitation spot. The density of this trap-free condensate is \textless \SI{1}{polariton\per\micro\meter\square}, reducing the phase noise induced by the interaction energy. Moreover, the back--flow effect, recently predicted for the non-parabolic polariton dispersion, is observed here for the first time in the fast-expanding wavepacket.
\end{abstract}


%\date{\today}
\maketitle

Under suitable conditions, light-matter interaction can be strong enough to drive the coherent exchange of energy between photons and electrons \cite{Hopfield1958}. This is the paradigm of microcavity exciton-polaritons: quasi-particles created by the strong coupling between the photonic mode of a microcavity and the excitonic transition of semiconductor quantum wells \cite{Weisbuch1992}. Polaritons manifest their composite nature with a combination of photonic and excitonic properties \cite{Dominici2014}. Thanks to their photonic component, polaritons can ballistically propagate in the plane of the microcavity with velocities up to a few percent of the speed of light \cite{Freixanet2000}. On the other hand, the exciton component results in strong optical nonlinearities and induces an energy renormalization of the polariton dispersion at high densities \cite{Senellart1999}. This energy shift can be much larger than the linewidth and is at the foundation of most polaritonic effects and applications \cite{Baas2004b, Amo2010, Ballarini2013}.
As bosonic quasiparticles, polaritons experience final-state stimulated scattering, which results, above a density threshold, in a laser-like emission without population inversion, a collective phenomenon that is explained in the framework of Bose-Einstein condensation \cite{Baumberg2000, Deng2002, Kasprzak2006}. A unique feature of polariton condensates is their driven/dissipative nature, in which the steady state is reached through a dynamical balance of pumping and dissipation.

Polariton condensates have been experimentally observed in different materials, both inorganic %(CdTe, GaAs, GaN, ZnO)
\cite{Richard2005, Balili2007, Christopoulos2007, Guillet2011} and organic semiconductors \cite{Daskalakis2014, Plumhof2014}, and thanks to their light mass, condensation can be achieved also at room temperature \cite{Baumberg2008}. However, differently from their atomic counterpart, these condensates suffer from dephasing and density fluctuations induced by the interactions with the exciton reservoir, effectively resulting in multimode condensates \cite{Kasprzak2008, Love2008, Kim2016}. The exciton reservoir, which is constantly feeding polaritons, also acts as a trapping mechanism, confining the condensation process within the region of the excitation spot \cite{Tassone1997, Wouters2008b, Roumpos2010, yama1}. Moreover, polariton condensation is often localized in potential minima caused by imperfections of the sample structure, yielding a fragmentation of the phase coherence \cite{Wouters2008, Baas2008, Lagoudakis2011, Thunert2016, Daskalakis2015}.
All these aspects of polariton condensates are not welcomed as they blur the fundamental character of the phenomenon by disrupting it with technical impediments. These become obstacles for fundamental studies and prospective applications with polaritons, such as investigating out-of-equilibrium phase transitions or to implement simulators and related devices \cite{Berloff2016}.
On the other hand, confinement of polaritons in one dimensional structures provides a striking evidence of the mechanism of expulsion and acceleration of polariton condensates far from the exciton reservoir \cite{Wertz2010, Wouters2010b, Wouters2012, Anton2013b}. In two-dimensional (2D) structures, interferences from scattering potentials, effects of laser-induced confinement and the wedge in the microcavity thickness, add additional difficulties in achieving extended and uniform condensates, even in samples with long polariton lifetimes \cite{Christmann2012, Cristofolini2013, Gao2016, Sun2016}.

In this work, we use a high quality 2D microcavity without spatial inhomogeneities, and observe the formation of a condensate that extends much beyond the laser spot region.
Photoluminescence measurements demonstrate the expansion of polaritons from the excitation spot and the subsequent relaxation into the lowest energy level at the bottom of the polariton dispersion. The extended condensate is formed thanks to two main ingredients: the high homogeneity of the sample, which avoids localization effects, and the long radiative lifetime ($\sim\SI{100}{\pico\second}$), that allows the propagating polariton to relax into the ground state. Moreover, to allow a uniform 2D expansion, the cavity wedge is almost absent in this sample, preventing a detuning-induced acceleration along preferential directions.  Remarkably, condensation occurs at low densities ($\sim\SI{0.1}{polariton\per\micro\meter\squared}$), without the presence of the exciton reservoir, and covers an area of more than $\SI{0.03}{\milli\meter\squared}$. A theoretical model that combines the hydrodynamics of the expanding polaritons with their energy relaxation reproduces the experimental energy-resolved spatial profiles. This confirms that, above a threshold power, phonon-mediated scattering into the lowest energy
mode is effective in forming an extended 2D polariton condensate.
Such a configuration allows us to observe a novel feature of the polariton flow, namely, a \emph{backflow} effect, that is a counterintuitive consequence of the polariton negative mass that makes polariton propagate in the direction opposite to their momentum,
thereby dragging them inwards, i.e., toward the exciting laser spot.

The sample used in this study is a high quality-factor $3/2$ $\lambda$ GaAs/AlGaAs planar cavity containing 12 GaAs quantum wells placed at three anti-node positions of the electric field. The front (back) mirror consists of 34 (40) pair of AlAs/Al$_{0.2}$Ga$_{0.8}$As layers. The Rabi energy splitting is \SI{16}{\milli\electronvolt} and the cavity-exciton detuning is of~$\SI{-2}{\milli\electronvolt}$. Experiments are performed under non-resonant excitation with a low-noise, narrow-linewidth Ti:sapphire laser with stabilized output frequency~\cite{note:note1}.
The sample emission is collected and imaged on the entrance slit of a streak camera coupled to a spectrometer in order to measure the time-, energy- and space-resolved polariton dynamics.

\begin{figure}[thbp]
 \centering
\includegraphics[width=0.48\textwidth]{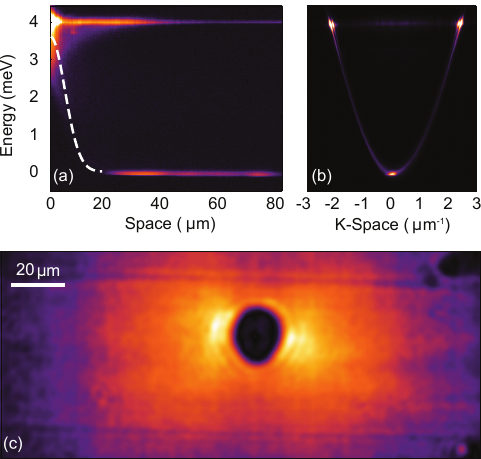}
\caption{\textbf{(a)} Energy versus real-space polariton emission along one direction, starting from the center of the pumping spot, displaying the expanding polaritons and the bottom condensate. The white-dashed line indicates the Gaussian potential induced by the excitonic repulsive interactions under the laser spot. \textbf{(b)} Energy-resolved reciprocal-space emission measured above the condensation threshold. The energy scale in the vertical axes is the same as in \textbf{(a)}. \textbf{(c)} Two-dimensional spatial emission map at $E=\SI{0.25}{\milli\electronvolt}$. Two random misfit dislocations, separated by about $\SI{80}{\micro\meter}$, appear as horizontal dark lines in the figure. 
}
\label{fig:1}
\end{figure}
Under nonresonant excitation, a high density of excitons accumulate within the region of the pumping spot, inducing a blueshift of the polariton energy proportional to their repulsive interaction strength. Outside the optically pumped area, the density of uncoupled excitons decreases quickly in space, due to the small exciton diffusion length (2--5 microns), and the polariton energy recovers the linear regime. The potential landscape, evidenced in Fig.~\ref{fig:1}(a) by a dashed-white line, reproduces the Gaussian profile of the exciting beam. In Fig.~\ref{fig:1}(a), the emitted intensity is energy- and spatially-resolved (vertical and horizontal axis, respectively) along one direction passing through the center of the excitation spot. The high-energy polaritons sitting at the top (\SI{4}{\milli\electronvolt} above the bottom energy) and formed at the center of the laser spot, expands radially outwards with a large in-plane wavevector ($k\sim\SI{2}{\per\micro\meter}$), as can be seen in the cross-section of the lower polariton dispersion (LPB, energy distribution in momentum space) shown in Fig.~\ref{fig:1}(b). The macroscopic occupation of the lowest energy mode ($k=0$), visible at the bottom of the LPB in Fig.~\ref{fig:1}(b), corresponds to the condensation outside of the spot region in Fig.~\ref{fig:1}(a). At the same time, also lower energy states along the whole dispersion ($k\leq\SI{2}{\per\micro\meter}$) are occupied and expand.
The high spatial homogeneity of the sample allows the polariton gas to expand uniformly, as shown in the two-dimensional, energy-filtered space map shown in Fig.~\ref{fig:1}(c).

In order to study the polariton dynamics, the steady state, populated through the continuous wave (CW) pump laser, is perturbed by focusing an additional \SI{100}{\femto\second} pulsed beam on top of the CW laser (both lasers are tuned to nonresonantly excite the system at the first minimum of the mirrors' stop band). 
The evolution of the additional polaritons injected by the pulse is recorded with a time resolution of \SI{5}{\pico\second}.
 \begin{figure}[thbp]
 \centering
\includegraphics[width=0.48\textwidth]{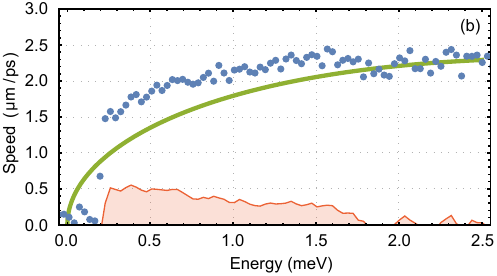}
\caption{The propagation speeds extracted by time-resolved measurements (blue dots) are compared to the polariton velocities as calculated from the LPB (green line). Faster expansions than expected are observed for energies below $E=\SI{1.7}{\milli\electronvolt}$ due to the prevailing effect of relaxation from higher energy state.}
\label{fig:2}
\end{figure} 
By extracting the space-time images at different energies, the expansion velocity can be estimated from the slope of the emission intensity~\cite{note:note2}. These velocities are compared in Fig.~\ref{fig:2} to the group velocities $v_{g}=\frac{1}{\hbar} \frac{\partial E_{LPB}(k)}{\partial k} $ calculated from the LPB dispersion and the difference is indicated by the red-filling region, showing that the effect of relaxation from higher energy states becomes considerable at lower energies.
To discard the possibility that the long propagation lengths can be explained by a faster expansion, for example due to a significant wedge in the cavity~\cite{Steger2013,Nelson2013}, we compare Fig.~\ref{fig:2} with a direct estimation of the polariton wavevectors. This is obtained by the optical tomography of a region of the sample where a single natural defect, injected in the DBR heterostructure during the growing process, breaks the spatial homogeneity of the surface. The pumping spot, the expanding fluid and the defect are clearly visible in Fig.~\ref{fig:3}(a), which shows the $(x,y)$ image of the emission at $E=\SI{0.5}{\milli\electronvolt}$. The region around the defect is magnified and filtered at different energies, showing increasing flow speeds from Fig.~\ref{fig:3}(b) to Fig.~\ref{fig:3}(d). The wavevectors extracted from the interference pattern (Fig.~\ref{fig:3}(f)) as $k=\frac{\pi}{\Delta x}$ match perfectly with the polariton dispersion, as shown in Fig.~\ref{fig:3}(e), also at the lowest energies. This shows unambiguously that the rate at which the polariton population distributes in space is the sum of two rates: the ballistic propagation and the filling rate due to relaxation processes. The second one becomes important for states close to the bottom of the LPB, and helps in reaching distances longer than expected from the bare propagation of low-speed polaritons.
\begin{figure}[t]
\centering
\includegraphics[width=0.48\textwidth]{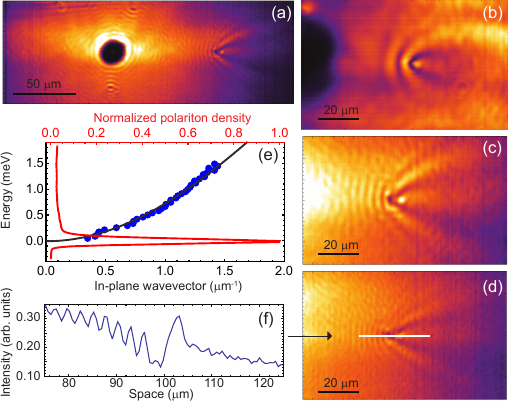}
\caption{
\textbf{(a)} Two-dimensional spatial emission at a given energy with propagation against a natural defect ($E=\SI{0.5}{\milli\electronvolt}$). \textbf{(b-d)} Particular cases of intensity oscillations against the defect with the characteristic angle of the defect shadow for different propagation velocities ($E$=\SI{0.1}, \SI{0.5}, \SI{1.1}{\milli\electronvolt} in panels \textbf{(b)}, \textbf{(c)} and \textbf{(d)}, respectively). \textbf{(e)} Energies versus wavevector extracted from the LPB (black line) and from the periodical spatial oscillations (blue dots). Polariton density as a function of energy is shown by the red line.
\textbf{(f)} Intensity cross section of the emission along the white line in panel \textbf{(d)}.
}
\label{fig:3}
\end{figure}
The dynamic of the fluid passing a defect is recorded here for a wide range of energies thanks to the continuum of states formed along the dispersion,  suggesting that, at higher densities, a new class of experiments on polariton superfluidity could be performed on expanding clouds free from reservoir artifacts and involving macroscopic distances.

\begin{figure}[t]
 \centering
\includegraphics[width=0.48\textwidth]{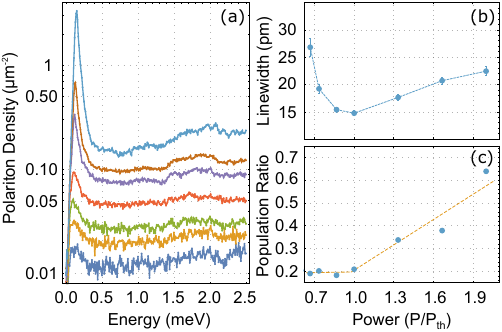}
\caption{
\textbf{(a)} Polariton density at a distance of \SI{40}{\micro\meter} from the excitation spot as a function of energy for the same pump powers as in \textbf{(b-c)}. The threshold power is $P_{th}=\SI{25}{\milli\watt}$. \textbf{(b)} Reciprocal-space linewidth as a function of power (same horizontal scale as in \textbf{(c)}). \textbf{(c)} Population ratio between the lowest energy state and the whole continuum of states at $\SI{40}{\micro\meter}$ from the spot region, showing the nonlinear increase above the condensation threshold. 
}
\label{fig:4}
\end{figure}

Remarkably, the polariton density outside of the pumped region, at the bottom of the LPB, manifests a nonlinear increase as a function of the pumping power for extremely low density values, as shown in Fig.~\ref{fig:4}(a)~\cite{Steger2016}. At the same time, the formation of the bottom condensate is marked by a narrowing of the linewidth, as shown in Fig.~\ref{fig:4}(b). The low density of the extended condensate diminishes the effects of polariton-polariton interaction, reducing the intrinsic dephasing of the condensate \cite{Kim2016} and making this configuration appealing both for applications and for future investigations of phase transition dynamics in polariton systems~\cite{Caputo2016}. In Fig.~\ref{fig:4}(c), the ratio between the density in the lowest energy state and the whole expanding cloud is compared at different excitation powers, showing a nonlinear increase at the condensation threshold. 

To reproduce the experimental observation, a theory that joins hydrodynamics and relaxation of an expanding and relaxing condensate is developed by combining the mean field description of the dynamics given by the Gross-Pitaevskii equation (GPE)~\cite{Wouters2008b,Carusotto13a} with the rate equations that account for stimulated scattering due to the interaction with a phonon bath~\cite{Doan2005,Wouters2010b}. %In order to obtain a simplified differential equation describing the steady-state polariton distribution in energy and space, we adapt the recent approach that merges both components of the dynamics at a level of the description that involves only the polariton density 
A simple differential equation that describes the steady-state polariton distribution in both energy and space can be derived with both components of the dynamics treated at the mean-field level~\cite{Bobrovska2016, note:note3}. 
The results of the calculation reproduce the observed results using the experimental parameters and confirm that an extended polariton condensate is formed when the phonon-mediated stimulated scattering is included ~\cite{note:note3}. 

The fast-expanding polariton fluid is therefore acting as a pure polaritonic reservoir for the large-area condensate at $k=0$. Its kinetic energy is provided by the polariton blueshift under the laser spot and can be finely controlled by tuning the excitation intensity. For higher powers, the group velocity of the expanding wavepacket increases until the blueshift reaches the energy corresponding to the inflection point of the polariton dispersion. This change of curvature, shown in Fig.~5 at $|k|=\SI{2.3}{\micro\meter^{-1}}$, formally corresponds to a change of sign in the diffusive mass, that becomes negative above this point. As recently predicted~\cite{Colas2016}, the presence of positive and negative masses can produce a counter-propagating flux within the expanding wavepacket.
In our experimental configuration, this means that a back--flow, directed towards the position of the exciting laser, would appear in the 2D density maps for velocities across the inflection point. This linear effect, caused by the peculiar curvature of the polariton
dispersion, is the result of the self-interference between different components of the expanding wavepacket. Thanks to the high degree of coherence sustained over large distances by the expanding wavepacket, we are able to clearly observe it in our sample.  The direct manifestation of the back--flow is indeed evident if we select the emission coming from a portion of space, and observe its density distribution in momentum space. With respect to Fig.~\ref{fig:3}(a), we now select only the emission coming from the half--space on the right of the pump spot. The measured polariton dispersion corresponding to this region is shown in Fig.~5(a-c) for three different excitation densities. If $|k|<k_{0}$, where $k$ is the expanding wavector and $k_{0}$ corresponds to the inflection point, only the portion of the dispersion associated with a current directed rightwards is populated ($k<0$). However, for higher blueshifts, also the opposite $k$-vectors, associated to a leftward propagation, become clearly visible ($k\sim k_{0}$). The effect can be directly observed also in real space thanks to the density modulations that appear \textit{before} the defect. The shadow cone that in Fig.~5(d) shows up only in the wake of the defect, is visible in Fig.~5(e) also in front of the defect, revealing the presence of a polariton flux in the opposite direction.

\begin{figure}[t]
 \centering
\includegraphics[width=0.48\textwidth]{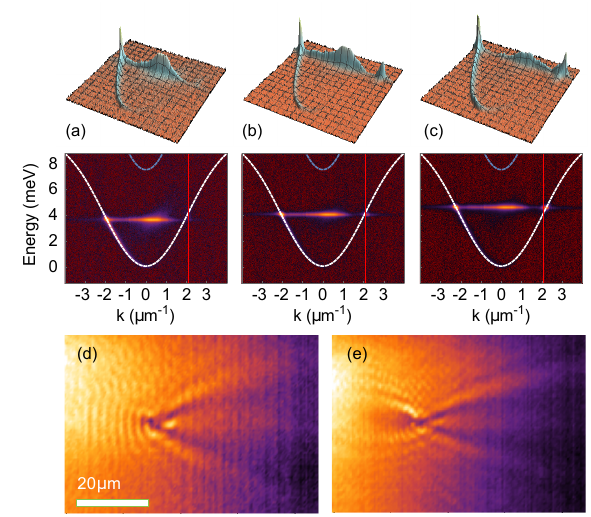}
\caption{\textbf{(a-c)} Density in momentum space of the flow directed horizontally to the right. The excitation power increases from \textbf{(a)} to \textbf{(c)}. At the inflection point, indicated by the red vertical line, the back--flow appears as a polariton density whose $k$-vector is directed to the left ($k>0$). \textbf{(d-e)} Real space density maps, taken as in Fig.~3, for expanding wavevectors below \textbf{(d)} and close \textbf{(e)} to the inflection point.}
\label{fig:5}
\end{figure}
In conclusion, the spontaneous formation of an extended 2D polariton condensate at the bottom of the LPB has been experimentally demonstrated. After non resonant excitation, the blueshifted region under the pumping spot injects an expanding flow of polaritons at the energy of the blue-shift, but also, through relaxation and expansion, a continuum of states along the dispersion. Two distinctive features of the presented experiments should be noted. The first is the backflow effect, observed within the expanding wavepacket thanks to the high surface homogeneity of the sample. The second is the relaxation until the bottom of the LPB, allowed by the long polariton lifetime. Above a threshold density, the condensate at the lowest energy forms at $k=0$, covering a spatial region larger then $\SI{0.03}{\milli\meter\squared}$ outside the pumping spot. These results are well reproduced by a simple theoretical model that captures the physics of expansion and relaxation of higher-energy polaritons. Our findings provide the closest realization so far of an infinite 2D polariton condensate, isolated from the exciton reservoir and with dilute, tunable densities. Such systems should allow to truly take advantage of microcavity polaritons for fundamental research on the dynamics of quantum fluids and driven/dissipative phase transitions.

\subsection*{Acknowledgments}
\begin{acknowledgments}
Funding from the POLAFLOW ERC Starting Grant is acknowledged.
\end{acknowledgments}

%%% Bibliography %%%%
%%%%%%%%%%%%%

\bibliographystyle{apsrev}
%\bibliography{Sci-degiorgi20140918}

\begin{thebibliography}{49}
\expandafter\ifx\csname natexlab\endcsname\relax\def\natexlab#1{#1}\fi
\expandafter\ifx\csname bibnamefont\endcsname\relax
  \def\bibnamefont#1{#1}\fi
\expandafter\ifx\csname bibfnamefont\endcsname\relax
  \def\bibfnamefont#1{#1}\fi
\expandafter\ifx\csname citenamefont\endcsname\relax
  \def\citenamefont#1{#1}\fi
\expandafter\ifx\csname url\endcsname\relax
  \def\url#1{\texttt{#1}}\fi
\expandafter\ifx\csname urlprefix\endcsname\relax\def\urlprefix{URL }\fi
\providecommand{\bibinfo}[2]{#2}
\providecommand{\eprint}[2][]{\url{#2}}

\bibitem[{\citenamefont{Hopfield}(1958)}]{Hopfield1958}
\bibinfo{author}{\bibfnamefont{J.~J.} \bibnamefont{Hopfield}},
  \bibinfo{journal}{Phys. Rev.} \textbf{\bibinfo{volume}{\textbf{112}}},
  \bibinfo{pages}{1555} (\bibinfo{year}{1958}).

\bibitem[{\citenamefont{Weisbuch et~al.}(1992)\citenamefont{Weisbuch, Nishioka,
  Ishikawa, and Arakawa}}]{Weisbuch1992}
\bibinfo{author}{\bibfnamefont{C.}~\bibnamefont{Weisbuch}},
  \bibinfo{author}{\bibfnamefont{M.}~\bibnamefont{Nishioka}},
  \bibinfo{author}{\bibfnamefont{A.}~\bibnamefont{Ishikawa}}, \bibnamefont{and}
  \bibinfo{author}{\bibfnamefont{Y.}~\bibnamefont{Arakawa}},
  \bibinfo{journal}{Phys. Rev. Lett.} \textbf{\bibinfo{volume}{\textbf{69}}},
  \bibinfo{pages}{3314} (\bibinfo{year}{1992}).

\bibitem[{\citenamefont{Dominici et~al.}(2014)\citenamefont{Dominici, Colas,
  Donati, Restrepo~Cuartas, De~Giorgi, Ballarini, Guirales, Lopez~Carreno,
  Bramati, Gigli et~al.}}]{Dominici2014}
\bibinfo{author}{\bibfnamefont{L.}~\bibnamefont{Dominici}},
  \bibinfo{author}{\bibfnamefont{D.}~\bibnamefont{Colas}},
  \bibinfo{author}{\bibfnamefont{S.}~\bibnamefont{Donati}},
  \bibinfo{author}{\bibfnamefont{J.}~\bibnamefont{Restrepo~Cuartas}},
  \bibinfo{author}{\bibfnamefont{M.}~\bibnamefont{De~Giorgi}},
  \bibinfo{author}{\bibfnamefont{D.}~\bibnamefont{Ballarini}},
  \bibinfo{author}{\bibfnamefont{G.}~\bibnamefont{Guirales}},
  \bibinfo{author}{\bibfnamefont{J.}~\bibnamefont{Lopez~Carreno}},
  \bibinfo{author}{\bibfnamefont{A.}~\bibnamefont{Bramati}},
  \bibinfo{author}{\bibfnamefont{G.}~\bibnamefont{Gigli}},
  \bibnamefont{et~al.}, \bibinfo{journal}{Phys. Rev. Lett.}
  \textbf{\bibinfo{volume}{113}}, \bibinfo{pages}{226401}
  (\bibinfo{year}{2014}).

\bibitem[{\citenamefont{Freixanet et~al.}(2000)\citenamefont{Freixanet,
  Sermage, Tiberj, and Planel}}]{Freixanet2000}
\bibinfo{author}{\bibfnamefont{T.}~\bibnamefont{Freixanet}},
  \bibinfo{author}{\bibfnamefont{B.}~\bibnamefont{Sermage}},
  \bibinfo{author}{\bibfnamefont{A.}~\bibnamefont{Tiberj}}, \bibnamefont{and}
  \bibinfo{author}{\bibfnamefont{R.}~\bibnamefont{Planel}},
  \bibinfo{journal}{Phys. Rev. B} \textbf{\bibinfo{volume}{\textbf{61}}},
  \bibinfo{pages}{7233} (\bibinfo{year}{2000}).

\bibitem[{\citenamefont{Senellart and Bloch}(1999)}]{Senellart1999}
\bibinfo{author}{\bibfnamefont{P.}~\bibnamefont{Senellart}} \bibnamefont{and}
  \bibinfo{author}{\bibfnamefont{J.}~\bibnamefont{Bloch}},
  \bibinfo{journal}{Phys. Rev. Lett.} \textbf{\bibinfo{volume}{82}},
  \bibinfo{pages}{1233} (\bibinfo{year}{1999}).

\bibitem[{\citenamefont{Baas et~al.}(2004)\citenamefont{Baas, Karr, Eleuch, and
  Giacobino}}]{Baas2004b}
\bibinfo{author}{\bibfnamefont{A.}~\bibnamefont{Baas}},
  \bibinfo{author}{\bibfnamefont{J.~P.} \bibnamefont{Karr}},
  \bibinfo{author}{\bibfnamefont{H.}~\bibnamefont{Eleuch}}, \bibnamefont{and}
  \bibinfo{author}{\bibfnamefont{E.}~\bibnamefont{Giacobino}},
  \bibinfo{journal}{Phys. Rev. A} \textbf{\bibinfo{volume}{69}},
  \bibinfo{pages}{023809} (\bibinfo{year}{2004}).

\bibitem[{\citenamefont{Amo et~al.}(2010)\citenamefont{Amo, Liew, Adrados,
  Houdre, Giacobino, Kavokin, and Bramati}}]{Amo2010}
\bibinfo{author}{\bibfnamefont{A.}~\bibnamefont{Amo}},
  \bibinfo{author}{\bibfnamefont{T.~C.~H.} \bibnamefont{Liew}},
  \bibinfo{author}{\bibfnamefont{C.}~\bibnamefont{Adrados}},
  \bibinfo{author}{\bibfnamefont{R.}~\bibnamefont{Houdre}},
  \bibinfo{author}{\bibfnamefont{E.}~\bibnamefont{Giacobino}},
  \bibinfo{author}{\bibfnamefont{A.~V.} \bibnamefont{Kavokin}},
  \bibnamefont{and} \bibinfo{author}{\bibfnamefont{A.}~\bibnamefont{Bramati}},
  \bibinfo{journal}{Nat. Photon.} \textbf{\bibinfo{volume}{4}},
  \bibinfo{pages}{361} (\bibinfo{year}{2010}).

\bibitem[{\citenamefont{Ballarini et~al.}(2013)\citenamefont{Ballarini,
  DeGiorgi, Cancellieri, Houdr\'e, Giacobino, Cingolani, Bramati, Gigli, and
  Sanvitto}}]{Ballarini2013}
\bibinfo{author}{\bibfnamefont{D.}~\bibnamefont{Ballarini}},
  \bibinfo{author}{\bibfnamefont{M.}~\bibnamefont{DeGiorgi}},
  \bibinfo{author}{\bibfnamefont{E.}~\bibnamefont{Cancellieri}},
  \bibinfo{author}{\bibfnamefont{R.}~\bibnamefont{Houdr\'e}},
  \bibinfo{author}{\bibfnamefont{E.}~\bibnamefont{Giacobino}},
  \bibinfo{author}{\bibfnamefont{R.}~\bibnamefont{Cingolani}},
  \bibinfo{author}{\bibfnamefont{A.}~\bibnamefont{Bramati}},
  \bibinfo{author}{\bibfnamefont{G.}~\bibnamefont{Gigli}}, \bibnamefont{and}
  \bibinfo{author}{\bibfnamefont{D.}~\bibnamefont{Sanvitto}},
  \bibinfo{journal}{Nat. Commun.} \textbf{\bibinfo{volume}{4}},
  \bibinfo{pages}{1778} (\bibinfo{year}{2013}).

\bibitem[{\citenamefont{Baumberg et~al.}(2000)\citenamefont{Baumberg, Savvidis,
  Stevenson, Tartakovskii, Skolnick, Whittaker, and Roberts}}]{Baumberg2000}
\bibinfo{author}{\bibfnamefont{J.~J.} \bibnamefont{Baumberg}},
  \bibinfo{author}{\bibfnamefont{P.~G.} \bibnamefont{Savvidis}},
  \bibinfo{author}{\bibfnamefont{R.~M.} \bibnamefont{Stevenson}},
  \bibinfo{author}{\bibfnamefont{A.~I.} \bibnamefont{Tartakovskii}},
  \bibinfo{author}{\bibfnamefont{M.~S.} \bibnamefont{Skolnick}},
  \bibinfo{author}{\bibfnamefont{D.~M.} \bibnamefont{Whittaker}},
  \bibnamefont{and} \bibinfo{author}{\bibfnamefont{J.~S.}
  \bibnamefont{Roberts}}, \bibinfo{journal}{Phys. Rev. B}
  \textbf{\bibinfo{volume}{\textbf{62}}}, \bibinfo{pages}{R16247}
  (\bibinfo{year}{2000}).

\bibitem[{\citenamefont{Deng et~al.}(2002)\citenamefont{Deng, Weihs, Santori,
  Bloch, and Yamamoto}}]{Deng2002}
\bibinfo{author}{\bibfnamefont{H.}~\bibnamefont{Deng}},
  \bibinfo{author}{\bibfnamefont{G.}~\bibnamefont{Weihs}},
  \bibinfo{author}{\bibfnamefont{C.}~\bibnamefont{Santori}},
  \bibinfo{author}{\bibfnamefont{J.}~\bibnamefont{Bloch}}, \bibnamefont{and}
  \bibinfo{author}{\bibfnamefont{Y.}~\bibnamefont{Yamamoto}},
  \bibinfo{journal}{Science} \textbf{\bibinfo{volume}{\textbf{298}}},
  \bibinfo{pages}{199} (\bibinfo{year}{2002}).

\bibitem[{\citenamefont{Kasprzak et~al.}(2006)\citenamefont{Kasprzak, Richard,
  Kundermann, Baas, Jeambrun, Keeling, Marchetti, Szymanska, Andre, Staehli
  et~al.}}]{Kasprzak2006}
\bibinfo{author}{\bibfnamefont{J.}~\bibnamefont{Kasprzak}},
  \bibinfo{author}{\bibfnamefont{M.}~\bibnamefont{Richard}},
  \bibinfo{author}{\bibfnamefont{S.}~\bibnamefont{Kundermann}},
  \bibinfo{author}{\bibfnamefont{A.}~\bibnamefont{Baas}},
  \bibinfo{author}{\bibfnamefont{P.}~\bibnamefont{Jeambrun}},
  \bibinfo{author}{\bibfnamefont{J.~M.~J.} \bibnamefont{Keeling}},
  \bibinfo{author}{\bibfnamefont{F.~M.} \bibnamefont{Marchetti}},
  \bibinfo{author}{\bibfnamefont{M.~H.} \bibnamefont{Szymanska}},
  \bibinfo{author}{\bibfnamefont{R.}~\bibnamefont{Andre}},
  \bibinfo{author}{\bibfnamefont{J.~L.} \bibnamefont{Staehli}},
  \bibnamefont{et~al.}, \bibinfo{journal}{Nature}
  \textbf{\bibinfo{volume}{443}}, \bibinfo{pages}{409} (\bibinfo{year}{2006}).

\bibitem[{\citenamefont{Richard et~al.}(2005)\citenamefont{Richard, Kasprzak,
  Andr\'{e}, Romestain, Dang, Malpuech, and Kavokin}}]{Richard2005}
\bibinfo{author}{\bibfnamefont{M.}~\bibnamefont{Richard}},
  \bibinfo{author}{\bibfnamefont{J.}~\bibnamefont{Kasprzak}},
  \bibinfo{author}{\bibfnamefont{R.}~\bibnamefont{Andr\'{e}}},
  \bibinfo{author}{\bibfnamefont{R.}~\bibnamefont{Romestain}},
  \bibinfo{author}{\bibfnamefont{L.~S.} \bibnamefont{Dang}},
  \bibinfo{author}{\bibfnamefont{G.}~\bibnamefont{Malpuech}}, \bibnamefont{and}
  \bibinfo{author}{\bibfnamefont{A.}~\bibnamefont{Kavokin}},
  \bibinfo{journal}{Phys. Rev. B} \textbf{\bibinfo{volume}{\textbf{72}}},
  \bibinfo{pages}{201301(R)} (\bibinfo{year}{2005}).

\bibitem[{\citenamefont{Balili et~al.}(2007)\citenamefont{Balili, Hartwell,
  Snoke, Pfeiffer, and West}}]{Balili2007}
\bibinfo{author}{\bibfnamefont{R.}~\bibnamefont{Balili}},
  \bibinfo{author}{\bibfnamefont{V.}~\bibnamefont{Hartwell}},
  \bibinfo{author}{\bibfnamefont{D.}~\bibnamefont{Snoke}},
  \bibinfo{author}{\bibfnamefont{L.}~\bibnamefont{Pfeiffer}}, \bibnamefont{and}
  \bibinfo{author}{\bibfnamefont{K.}~\bibnamefont{West}},
  \bibinfo{journal}{Science} \textbf{\bibinfo{volume}{\textbf{316}}},
  \bibinfo{pages}{1007} (\bibinfo{year}{2007}).

\bibitem[{\citenamefont{Christopoulos et~al.}(2007)\citenamefont{Christopoulos,
  Baldassarri Hoger~von Hogersthal, Grundy, Lagoudakis, Kavokin, Baumberg,
  Christmann, Butte, Feltin, Carlin et~al.}}]{Christopoulos2007}
\bibinfo{author}{\bibfnamefont{S.}~\bibnamefont{Christopoulos}},
  \bibinfo{author}{\bibfnamefont{G.}~\bibnamefont{Baldassarri Hoger~von
  Hogersthal}}, \bibinfo{author}{\bibfnamefont{A.~J.~D.} \bibnamefont{Grundy}},
  \bibinfo{author}{\bibfnamefont{P.~G.} \bibnamefont{Lagoudakis}},
  \bibinfo{author}{\bibfnamefont{A.~V.} \bibnamefont{Kavokin}},
  \bibinfo{author}{\bibfnamefont{J.~J.} \bibnamefont{Baumberg}},
  \bibinfo{author}{\bibfnamefont{G.}~\bibnamefont{Christmann}},
  \bibinfo{author}{\bibfnamefont{R.}~\bibnamefont{Butte}},
  \bibinfo{author}{\bibfnamefont{E.}~\bibnamefont{Feltin}},
  \bibinfo{author}{\bibfnamefont{J.-F.} \bibnamefont{Carlin}},
  \bibnamefont{et~al.}, \bibinfo{journal}{Phys. Rev. Lett.}
  \textbf{\bibinfo{volume}{\textbf{98}}}, \bibinfo{pages}{126405}
  (\bibinfo{year}{2007}).

\bibitem[{\citenamefont{Guillet et~al.}(2011)\citenamefont{Guillet, Mexis,
  Levrat, Rossbach, Brimont, Bretagnon, Gil, Butt\'e, Grandjean, Orosz
  et~al.}}]{Guillet2011}
\bibinfo{author}{\bibfnamefont{T.}~\bibnamefont{Guillet}},
  \bibinfo{author}{\bibfnamefont{M.}~\bibnamefont{Mexis}},
  \bibinfo{author}{\bibfnamefont{J.}~\bibnamefont{Levrat}},
  \bibinfo{author}{\bibfnamefont{G.}~\bibnamefont{Rossbach}},
  \bibinfo{author}{\bibfnamefont{C.}~\bibnamefont{Brimont}},
  \bibinfo{author}{\bibfnamefont{T.}~\bibnamefont{Bretagnon}},
  \bibinfo{author}{\bibfnamefont{B.}~\bibnamefont{Gil}},
  \bibinfo{author}{\bibfnamefont{R.}~\bibnamefont{Butt\'e}},
  \bibinfo{author}{\bibfnamefont{N.}~\bibnamefont{Grandjean}},
  \bibinfo{author}{\bibfnamefont{L.}~\bibnamefont{Orosz}},
  \bibnamefont{et~al.}, \bibinfo{journal}{Appl. Phys. Lett.}
  \textbf{\bibinfo{volume}{99}}, \bibinfo{pages}{161104}
  (\bibinfo{year}{2011}).

\bibitem[{\citenamefont{Daskalakis et~al.}(2014)\citenamefont{Daskalakis,
  Maier, Murray, and K{\'e}na-Cohen}}]{Daskalakis2014}
\bibinfo{author}{\bibfnamefont{K.~S.} \bibnamefont{Daskalakis}},
  \bibinfo{author}{\bibfnamefont{S.~A.} \bibnamefont{Maier}},
  \bibinfo{author}{\bibfnamefont{R.}~\bibnamefont{Murray}}, \bibnamefont{and}
  \bibinfo{author}{\bibfnamefont{S.}~\bibnamefont{K{\'e}na-Cohen}},
  \bibinfo{journal}{Nat Mater} \textbf{\bibinfo{volume}{13}},
  \bibinfo{pages}{271} (\bibinfo{year}{2014}).

\bibitem[{\citenamefont{Plumhof et~al.}(2014)\citenamefont{Plumhof, Stoferle,
  Mai, Scherf, and Mahrt}}]{Plumhof2014}
\bibinfo{author}{\bibfnamefont{J.}~\bibnamefont{Plumhof}},
  \bibinfo{author}{\bibfnamefont{T.}~\bibnamefont{Stoferle}},
  \bibinfo{author}{\bibfnamefont{L.}~\bibnamefont{Mai}},
  \bibinfo{author}{\bibfnamefont{U.}~\bibnamefont{Scherf}}, \bibnamefont{and}
  \bibinfo{author}{\bibfnamefont{R.~F.} \bibnamefont{Mahrt}},
  \bibinfo{journal}{Nat. Mater.} \textbf{\bibinfo{volume}{13}},
  \bibinfo{pages}{247} (\bibinfo{year}{2014}).

\bibitem[{\citenamefont{Baumberg et~al.}(2008)\citenamefont{Baumberg, Kavokin,
  Christopoulos, Grundy, Butt\'e, Christmann, Solnyshkov, Malpuech, Baldassarri
  H\"oger~von H\"ogersthal, Feltin et~al.}}]{Baumberg2008}
\bibinfo{author}{\bibfnamefont{J.~J.} \bibnamefont{Baumberg}},
  \bibinfo{author}{\bibfnamefont{A.~V.} \bibnamefont{Kavokin}},
  \bibinfo{author}{\bibfnamefont{S.}~\bibnamefont{Christopoulos}},
  \bibinfo{author}{\bibfnamefont{A.~J.~D.} \bibnamefont{Grundy}},
  \bibinfo{author}{\bibfnamefont{R.}~\bibnamefont{Butt\'e}},
  \bibinfo{author}{\bibfnamefont{G.}~\bibnamefont{Christmann}},
  \bibinfo{author}{\bibfnamefont{D.~D.} \bibnamefont{Solnyshkov}},
  \bibinfo{author}{\bibfnamefont{G.}~\bibnamefont{Malpuech}},
  \bibinfo{author}{\bibfnamefont{G.}~\bibnamefont{Baldassarri H\"oger~von
  H\"ogersthal}}, \bibinfo{author}{\bibfnamefont{E.}~\bibnamefont{Feltin}},
  \bibnamefont{et~al.}, \bibinfo{journal}{Phys. Rev. Lett.}
  \textbf{\bibinfo{volume}{101}}, \bibinfo{pages}{136409}
  (\bibinfo{year}{2008}).

\bibitem[{\citenamefont{Kasprzak et~al.}(2008)\citenamefont{Kasprzak, Richard,
  Baas, Deveaud, Andre, Poizat, and Dang}}]{Kasprzak2008}
\bibinfo{author}{\bibfnamefont{J.}~\bibnamefont{Kasprzak}},
  \bibinfo{author}{\bibfnamefont{M.}~\bibnamefont{Richard}},
  \bibinfo{author}{\bibfnamefont{A.}~\bibnamefont{Baas}},
  \bibinfo{author}{\bibfnamefont{B.}~\bibnamefont{Deveaud}},
  \bibinfo{author}{\bibfnamefont{R.}~\bibnamefont{Andre}},
  \bibinfo{author}{\bibfnamefont{J.~P.}~\bibnamefont{Poizat}}, \bibnamefont{and}
  \bibinfo{author}{\bibfnamefont{L.~S.} \bibnamefont{Dang}},
  \bibinfo{journal}{Phys. Rev. Lett.} \textbf{\bibinfo{volume}{100}},
  \bibinfo{pages}{067402} (\bibinfo{year}{2008}).

\bibitem[{\citenamefont{Love et~al.}(2008)\citenamefont{Love, Krizhanovskii,
  Whittaker, Bouchekioua, Sanvitto, Rizeiqi, Bradley, Skolnick, Eastham, Andre
  et~al.}}]{Love2008}
\bibinfo{author}{\bibfnamefont{A.~P.~D.} \bibnamefont{Love}},
  \bibinfo{author}{\bibfnamefont{D.~N.} \bibnamefont{Krizhanovskii}},
  \bibinfo{author}{\bibfnamefont{D.~M.} \bibnamefont{Whittaker}},
  \bibinfo{author}{\bibfnamefont{R.}~\bibnamefont{Bouchekioua}},
  \bibinfo{author}{\bibfnamefont{D.}~\bibnamefont{Sanvitto}},
  \bibinfo{author}{\bibfnamefont{S.~A.} \bibnamefont{Rizeiqi}},
  \bibinfo{author}{\bibfnamefont{R.}~\bibnamefont{Bradley}},
  \bibinfo{author}{\bibfnamefont{M.~S.} \bibnamefont{Skolnick}},
  \bibinfo{author}{\bibfnamefont{P.~R.} \bibnamefont{Eastham}},
  \bibinfo{author}{\bibfnamefont{R.}~\bibnamefont{Andre}},
  \bibnamefont{et~al.}, \bibinfo{journal}{Phys. Rev. Lett.}
  \textbf{\bibinfo{volume}{101}}, \bibinfo{pages}{067404}
  (\bibinfo{year}{2008}).

\bibitem[{\citenamefont{Kim et~al.}(2016)\citenamefont{Kim, Zhang, Wang,
  Fischer, Brodbeck, Kamp, Schneider, H\"ofling, and Deng}}]{Kim2016}
\bibinfo{author}{\bibfnamefont{S.}~\bibnamefont{Kim}},
  \bibinfo{author}{\bibfnamefont{B.}~\bibnamefont{Zhang}},
  \bibinfo{author}{\bibfnamefont{Z.}~\bibnamefont{Wang}},
  \bibinfo{author}{\bibfnamefont{J.}~\bibnamefont{Fischer}},
  \bibinfo{author}{\bibfnamefont{S.}~\bibnamefont{Brodbeck}},
  \bibinfo{author}{\bibfnamefont{M.}~\bibnamefont{Kamp}},
  \bibinfo{author}{\bibfnamefont{C.}~\bibnamefont{Schneider}},
  \bibinfo{author}{\bibfnamefont{S.}~\bibnamefont{H\"ofling}},
  \bibnamefont{and} \bibinfo{author}{\bibfnamefont{H.}~\bibnamefont{Deng}},
  \bibinfo{journal}{Phys. Rev. X} \textbf{\bibinfo{volume}{6}},
  \bibinfo{pages}{011026} (\bibinfo{year}{2016}).

\bibitem[{\citenamefont{Tassone et~al.}(1997)\citenamefont{Tassone,
  Piermarocchi, Savona, Quattropani, and Schwendimann}}]{Tassone1997}
\bibinfo{author}{\bibfnamefont{F.}~\bibnamefont{Tassone}},
  \bibinfo{author}{\bibfnamefont{C.}~\bibnamefont{Piermarocchi}},
  \bibinfo{author}{\bibfnamefont{V.}~\bibnamefont{Savona}},
  \bibinfo{author}{\bibfnamefont{A.}~\bibnamefont{Quattropani}},
  \bibnamefont{and}
  \bibinfo{author}{\bibfnamefont{P.}~\bibnamefont{Schwendimann}},
  \bibinfo{journal}{Phys. Rev. B} \textbf{\bibinfo{volume}{56}},
  \bibinfo{pages}{7554} (\bibinfo{year}{1997}).

\bibitem[{\citenamefont{Wouters et~al.}(2008)\citenamefont{Wouters, Carusotto,
  and Ciuti}}]{Wouters2008b}
\bibinfo{author}{\bibfnamefont{M.}~\bibnamefont{Wouters}},
  \bibinfo{author}{\bibfnamefont{I.}~\bibnamefont{Carusotto}},
  \bibnamefont{and} \bibinfo{author}{\bibfnamefont{C.}~\bibnamefont{Ciuti}},
  \bibinfo{journal}{Phys. Rev. B} \textbf{\bibinfo{volume}{77}},
  \bibinfo{pages}{115340} (\bibinfo{year}{2008}).

\bibitem[{\citenamefont{Roumpos et~al.}(2010)\citenamefont{Roumpos, Nitsche,
  H\"ofling, Forchel, and Yamamoto}}]{Roumpos2010}
\bibinfo{author}{\bibfnamefont{G.}~\bibnamefont{Roumpos}},
  \bibinfo{author}{\bibfnamefont{W.~H.} \bibnamefont{Nitsche}},
  \bibinfo{author}{\bibfnamefont{S.}~\bibnamefont{H\"ofling}},
  \bibinfo{author}{\bibfnamefont{A.}~\bibnamefont{Forchel}}, \bibnamefont{and}
  \bibinfo{author}{\bibfnamefont{Y.}~\bibnamefont{Yamamoto}},
  \bibinfo{journal}{Phys. Rev. Lett.} \textbf{\bibinfo{volume}{104}},
  \bibinfo{pages}{126403} (\bibinfo{year}{2010}).

\bibitem[{\citenamefont{Roumpos et~al.}(2012)\citenamefont{Roumpos, Lohse,
  Nitsche, Keeling, Szyma{\'{n}}ska, Littlewood, L?ffler, H?fling, Worschech,
  Forchel et~al.}}]{yama1}
\bibinfo{author}{\bibfnamefont{G.}~\bibnamefont{Roumpos}},
  \bibinfo{author}{\bibfnamefont{M.}~\bibnamefont{Lohse}},
  \bibinfo{author}{\bibfnamefont{W.~H.} \bibnamefont{Nitsche}},
  \bibinfo{author}{\bibfnamefont{J.}~\bibnamefont{Keeling}},
  \bibinfo{author}{\bibfnamefont{M.~H.} \bibnamefont{Szyma{\'{n}}ska}},
  \bibinfo{author}{\bibfnamefont{P.~B.} \bibnamefont{Littlewood}},
  \bibinfo{author}{\bibfnamefont{A.}~\bibnamefont{L{\"{o}}ffler}},
  \bibinfo{author}{\bibfnamefont{S.}~\bibnamefont{H{\"{o}}fling}},
  \bibinfo{author}{\bibfnamefont{L.}~\bibnamefont{Worschech}},
  \bibinfo{author}{\bibfnamefont{A.}~\bibnamefont{Forchel}},
  \bibnamefont{et~al.}, \bibinfo{journal}{PNAS} \textbf{\bibinfo{volume}{109}},
  \bibinfo{pages}{6467} (\bibinfo{year}{2012}).

\bibitem[{\citenamefont{Wouters and Carusotto}(2008)}]{Wouters2008}
\bibinfo{author}{\bibfnamefont{M.}~\bibnamefont{Wouters}} \bibnamefont{and}
  \bibinfo{author}{\bibfnamefont{I.}~\bibnamefont{Carusotto}},
  \bibinfo{journal}{Superlattices and Microstructures}
  \textbf{\bibinfo{volume}{\textbf{43}}}, \bibinfo{pages}{524}
  (\bibinfo{year}{2008}).

\bibitem[{\citenamefont{Baas et~al.}(2008)\citenamefont{Baas, Lagoudakis,
  Richard, Andre, Dang, and Deveaud-Pledran}}]{Baas2008}
\bibinfo{author}{\bibfnamefont{A.}~\bibnamefont{Baas}},
  \bibinfo{author}{\bibfnamefont{K.~G.} \bibnamefont{Lagoudakis}},
  \bibinfo{author}{\bibfnamefont{M.}~\bibnamefont{Richard}},
  \bibinfo{author}{\bibfnamefont{R.}~\bibnamefont{Andre}},
  \bibinfo{author}{\bibfnamefont{L.~S.} \bibnamefont{Dang}}, \bibnamefont{and}
  \bibinfo{author}{\bibfnamefont{B.}~\bibnamefont{Deveaud-Pledran}},
  \bibinfo{journal}{Phys. Rev. Lett.} \textbf{\bibinfo{volume}{100}},
  \bibinfo{pages}{170401} (\bibinfo{year}{2008}).

\bibitem[{\citenamefont{Lagoudakis et~al.}(2011)\citenamefont{Lagoudakis,
  Manni, Pietka, Wouters, Liew, Savona, Kavokin, Andr??, and
  Deveaud-Pledran}}]{Lagoudakis2011}
\bibinfo{author}{\bibfnamefont{K.~G.} \bibnamefont{Lagoudakis}},
  \bibinfo{author}{\bibfnamefont{F.}~\bibnamefont{Manni}},
  \bibinfo{author}{\bibfnamefont{B.}~\bibnamefont{Pietka}},
  \bibinfo{author}{\bibfnamefont{M.}~\bibnamefont{Wouters}},
  \bibinfo{author}{\bibfnamefont{T.~C.~H.} \bibnamefont{Liew}},
  \bibinfo{author}{\bibfnamefont{V.}~\bibnamefont{Savona}},
  \bibinfo{author}{\bibfnamefont{A.~V.} \bibnamefont{Kavokin}},
  \bibinfo{author}{\bibfnamefont{R.}~\bibnamefont{Andr\'e}}, \bibnamefont{and}
  \bibinfo{author}{\bibfnamefont{B.}~\bibnamefont{Deveaud-Pledran}},
  \bibinfo{journal}{Phys. Rev. Lett.} \textbf{\bibinfo{volume}{106}},
  \bibinfo{pages}{115301} (\bibinfo{year}{2011}).

\bibitem[{\citenamefont{Thunert et~al.}(2016)\citenamefont{Thunert, Janot,
  Franke, Sturm, Michalsky, Mart\'in, Vi\~na, Rosenow, Grundmann, and
  Schmidt-Grund}}]{Thunert2016}
\bibinfo{author}{\bibfnamefont{M.}~\bibnamefont{Thunert}},
  \bibinfo{author}{\bibfnamefont{A.}~\bibnamefont{Janot}},
  \bibinfo{author}{\bibfnamefont{H.}~\bibnamefont{Franke}},
  \bibinfo{author}{\bibfnamefont{C.}~\bibnamefont{Sturm}},
  \bibinfo{author}{\bibfnamefont{T.}~\bibnamefont{Michalsky}},
  \bibinfo{author}{\bibfnamefont{M.~D.}~\bibnamefont{Mart\'in}},
  \bibinfo{author}{\bibfnamefont{L.}~\bibnamefont{Vi\~na}},
  \bibinfo{author}{\bibfnamefont{B.}~\bibnamefont{Rosenow}},
  \bibinfo{author}{\bibfnamefont{M.}~\bibnamefont{Grundmann}},
  \bibnamefont{and}
  \bibinfo{author}{\bibfnamefont{R.}~\bibnamefont{Schmidt-Grund}},
  \bibinfo{journal}{Phys. Rev. B} \textbf{\bibinfo{volume}{93}},
  \bibinfo{pages}{064203} (\bibinfo{year}{2016}).

\bibitem[{\citenamefont{Daskalakis et~al.}(2015)\citenamefont{Daskalakis,
  Maier, and K\'ena-Cohen}}]{Daskalakis2015}
\bibinfo{author}{\bibfnamefont{K.~S.} \bibnamefont{Daskalakis}},
  \bibinfo{author}{\bibfnamefont{S.~A.} \bibnamefont{Maier}}, \bibnamefont{and}
  \bibinfo{author}{\bibfnamefont{S.}~\bibnamefont{K\'ena-Cohen}},
  \bibinfo{journal}{Phys. Rev. Lett.} \textbf{\bibinfo{volume}{115}},
  \bibinfo{pages}{035301} (\bibinfo{year}{2015}).

\bibitem[{\citenamefont{Berloff et~al.}(2016)\citenamefont{Berloff, Kalinin,
  Silva, Langbein, and Lagoudakis}}]{Berloff2016}
\bibinfo{author}{\bibfnamefont{N.~G.} \bibnamefont{Berloff}},
  \bibinfo{author}{\bibfnamefont{K.}~\bibnamefont{Kalinin}},
  \bibinfo{author}{\bibfnamefont{M.}~\bibnamefont{Silva}},
  \bibinfo{author}{\bibfnamefont{W.}~\bibnamefont{Langbein}}, \bibnamefont{and}
  \bibinfo{author}{\bibfnamefont{P.~G.} \bibnamefont{Lagoudakis}},
  \bibinfo{journal}{arXiv:1607.06065v1 [cond-mat.mes-hall]}
  (\bibinfo{year}{2016}).

\bibitem[{\citenamefont{Wertz et~al.}(2010)\citenamefont{Wertz, Ferrier,
  Solnyshkov, Johne, Sanvitto, Lemaitre, Sagnes, Grousson, Kavokin, Senellart
  et~al.}}]{Wertz2010}
\bibinfo{author}{\bibfnamefont{E.}~\bibnamefont{Wertz}},
  \bibinfo{author}{\bibfnamefont{L.}~\bibnamefont{Ferrier}},
  \bibinfo{author}{\bibfnamefont{D.~D.} \bibnamefont{Solnyshkov}},
  \bibinfo{author}{\bibfnamefont{R.}~\bibnamefont{Johne}},
  \bibinfo{author}{\bibfnamefont{D.}~\bibnamefont{Sanvitto}},
  \bibinfo{author}{\bibfnamefont{A.}~\bibnamefont{Lemaitre}},
  \bibinfo{author}{\bibfnamefont{I.}~\bibnamefont{Sagnes}},
  \bibinfo{author}{\bibfnamefont{R.}~\bibnamefont{Grousson}},
  \bibinfo{author}{\bibfnamefont{A.~V.} \bibnamefont{Kavokin}},
  \bibinfo{author}{\bibfnamefont{P.}~\bibnamefont{Senellart}},
  \bibnamefont{et~al.}, \bibinfo{journal}{Nature Phys.}
  \textbf{\bibinfo{volume}{6}}, \bibinfo{pages}{860} (\bibinfo{year}{2010}).

\bibitem[{\citenamefont{Wouters et~al.}(2010)\citenamefont{Wouters, Liew, and
  Savona}}]{Wouters2010b}
\bibinfo{author}{\bibfnamefont{M.}~\bibnamefont{Wouters}},
  \bibinfo{author}{\bibfnamefont{T.~C.~H.} \bibnamefont{Liew}},
  \bibnamefont{and} \bibinfo{author}{\bibfnamefont{V.}~\bibnamefont{Savona}},
  \bibinfo{journal}{Phys. Rev. B} \textbf{\bibinfo{volume}{82}},
  \bibinfo{pages}{245315} (\bibinfo{year}{2010}).

\bibitem[{\citenamefont{Wouters}(2012)}]{Wouters2012}
\bibinfo{author}{\bibfnamefont{M.}~\bibnamefont{Wouters}},
  \bibinfo{journal}{New Journal of Physics} \textbf{\bibinfo{volume}{14}},
  \bibinfo{pages}{1} (\bibinfo{year}{2012}).

\bibitem[{\citenamefont{Anton et~al.}(2013)\citenamefont{Anton, Liew, Tosi,
  Martin, Gao, Hatzopoulos, Eldridge, Savvidis, and Vina}}]{Anton2013b}
\bibinfo{author}{\bibfnamefont{C.}~\bibnamefont{Anton}},
  \bibinfo{author}{\bibfnamefont{T.~C.~H.} \bibnamefont{Liew}},
  \bibinfo{author}{\bibfnamefont{G.}~\bibnamefont{Tosi}},
  \bibinfo{author}{\bibfnamefont{M.~D.}~\bibnamefont{Martin}},
  \bibinfo{author}{\bibfnamefont{T.}~\bibnamefont{Gao}},
  \bibinfo{author}{\bibfnamefont{Z.}~\bibnamefont{Hatzopoulos}},
  \bibinfo{author}{\bibfnamefont{P.~S.} \bibnamefont{Eldridge}},
  \bibinfo{author}{\bibfnamefont{P.~G.} \bibnamefont{Savvidis}},
  \bibnamefont{and} \bibinfo{author}{\bibfnamefont{L.}~\bibnamefont{Vina}},
  \bibinfo{journal}{Phys. Rev. B} \textbf{\bibinfo{volume}{88}},
  \bibinfo{pages}{035313} (\bibinfo{year}{2013}).

\bibitem[{\citenamefont{Christmann et~al.}(2012)\citenamefont{Christmann, Tosi,
  Berloff, Tsotsis, Eldridge, Hatzopoulos, Savvidis, and
  Baumberg}}]{Christmann2012}
\bibinfo{author}{\bibfnamefont{G.}~\bibnamefont{Christmann}},
  \bibinfo{author}{\bibfnamefont{G.}~\bibnamefont{Tosi}},
  \bibinfo{author}{\bibfnamefont{N.~G.} \bibnamefont{Berloff}},
  \bibinfo{author}{\bibfnamefont{P.}~\bibnamefont{Tsotsis}},
  \bibinfo{author}{\bibfnamefont{P.~S.} \bibnamefont{Eldridge}},
  \bibinfo{author}{\bibfnamefont{Z.}~\bibnamefont{Hatzopoulos}},
  \bibinfo{author}{\bibfnamefont{P.~G.} \bibnamefont{Savvidis}},
  \bibnamefont{and} \bibinfo{author}{\bibfnamefont{J.~J.}
  \bibnamefont{Baumberg}}, \bibinfo{journal}{Phys. Rev. B}
  \textbf{\bibinfo{volume}{85}}, \bibinfo{pages}{235303}
  (\bibinfo{year}{2012}).

\bibitem[{\citenamefont{Cristofolini et~al.}(2013)\citenamefont{Cristofolini,
  Dreismann, Christmann, Franchetti, Berloff, Tsotsis, Hatzopoulos, Savvidis,
  and Baumberg}}]{Cristofolini2013}
\bibinfo{author}{\bibfnamefont{P.}~\bibnamefont{Cristofolini}},
  \bibinfo{author}{\bibfnamefont{A.}~\bibnamefont{Dreismann}},
  \bibinfo{author}{\bibfnamefont{G.}~\bibnamefont{Christmann}},
  \bibinfo{author}{\bibfnamefont{G.}~\bibnamefont{Franchetti}},
  \bibinfo{author}{\bibfnamefont{N.~G.} \bibnamefont{Berloff}},
  \bibinfo{author}{\bibfnamefont{P.}~\bibnamefont{Tsotsis}},
  \bibinfo{author}{\bibfnamefont{Z.}~\bibnamefont{Hatzopoulos}},
  \bibinfo{author}{\bibfnamefont{P.~G.} \bibnamefont{Savvidis}},
  \bibnamefont{and} \bibinfo{author}{\bibfnamefont{J.~J.}
  \bibnamefont{Baumberg}}, \bibinfo{journal}{Phys. Rev. Lett.}
  \textbf{\bibinfo{volume}{110}}, \bibinfo{pages}{186403}
  (\bibinfo{year}{2013}).

\bibitem[{\citenamefont{Gao et~al.}(2016)\citenamefont{Gao, Estrecho, Li,
  Egorov, Ma, Winkler, Kamp, Schneider, H\"ofling, Truscott et~al.}}]{Gao2016}
\bibinfo{author}{\bibfnamefont{T.}~\bibnamefont{Gao}},
  \bibinfo{author}{\bibfnamefont{E.}~\bibnamefont{Estrecho}},
  \bibinfo{author}{\bibfnamefont{G.}~\bibnamefont{Li}},
  \bibinfo{author}{\bibfnamefont{O.~A.} \bibnamefont{Egorov}},
  \bibinfo{author}{\bibfnamefont{X.}~\bibnamefont{Ma}},
  \bibinfo{author}{\bibfnamefont{K.}~\bibnamefont{Winkler}},
  \bibinfo{author}{\bibfnamefont{M.}~\bibnamefont{Kamp}},
  \bibinfo{author}{\bibfnamefont{C.}~\bibnamefont{Schneider}},
  \bibinfo{author}{\bibfnamefont{S.}~\bibnamefont{H\"ofling}},
  \bibinfo{author}{\bibfnamefont{A.~G.} \bibnamefont{Truscott}},
  \bibnamefont{et~al.}, \bibinfo{journal}{Phys. Rev. Lett.}
  \textbf{\bibinfo{volume}{117}}, \bibinfo{pages}{097403}
  (\bibinfo{year}{2016}).

\bibitem[{\citenamefont{Sun et~al.}(2017)\citenamefont{Sun, Wen, Yoon, Liu,
  Steger, Pfeiffer, West, Snoke, and Nelson}}]{Sun2016}
\bibinfo{author}{\bibfnamefont{Y.}~\bibnamefont{Sun}},
  \bibinfo{author}{\bibfnamefont{P.}~\bibnamefont{Wen}},
  \bibinfo{author}{\bibfnamefont{Y.}~\bibnamefont{Yoon}},
  \bibinfo{author}{\bibfnamefont{G.}~\bibnamefont{Liu}},
  \bibinfo{author}{\bibfnamefont{M.}~\bibnamefont{Steger}},
  \bibinfo{author}{\bibfnamefont{L.~N.} \bibnamefont{Pfeiffer}},
  \bibinfo{author}{\bibfnamefont{K.}~\bibnamefont{West}},
  \bibinfo{author}{\bibfnamefont{D.~W.} \bibnamefont{Snoke}}, \bibnamefont{and}
  \bibinfo{author}{\bibfnamefont{K.~A.} \bibnamefont{Nelson}},
  \bibinfo{journal}{Phys. Rev. Lett.} \textbf{\bibinfo{volume}{118}},
  \bibinfo{pages}{016602} (\bibinfo{year}{2017}).

\bibitem[{not({\natexlab{a}})}]{note:note1}
\bibinfo{note}{See details on the excitation scheme in the Supplemental Material}.

\bibitem[{not({\natexlab{b}})}]{note:note2}
\bibinfo{note}{See Supplemental Material for the whole set of data (movie) and
  Fig. S2 and Fig. S3 for further details}.

\bibitem[{\citenamefont{Steger et~al.}(2013)\citenamefont{Steger, Liu, Nelsen,
  Gautham, Snoke, Balili, Pfeiffer, and West}}]{Steger2013}
\bibinfo{author}{\bibfnamefont{M.}~\bibnamefont{Steger}},
  \bibinfo{author}{\bibfnamefont{G.}~\bibnamefont{Liu}},
  \bibinfo{author}{\bibfnamefont{B.}~\bibnamefont{Nelsen}},
  \bibinfo{author}{\bibfnamefont{C.}~\bibnamefont{Gautham}},
  \bibinfo{author}{\bibfnamefont{D.~W.} \bibnamefont{Snoke}},
  \bibinfo{author}{\bibfnamefont{R.}~\bibnamefont{Balili}},
  \bibinfo{author}{\bibfnamefont{L.}~\bibnamefont{Pfeiffer}}, \bibnamefont{and}
  \bibinfo{author}{\bibfnamefont{K.}~\bibnamefont{West}},
  \bibinfo{journal}{Phys. Rev. B} \textbf{\bibinfo{volume}{88}},
  \bibinfo{pages}{235314} (\bibinfo{year}{2013}).

\bibitem[{\citenamefont{Nelsen et~al.}(2013)\citenamefont{Nelsen, Liu, Steger,
  Snoke, Balili, West, and Pfeiffer}}]{Nelson2013}
\bibinfo{author}{\bibfnamefont{B.}~\bibnamefont{Nelsen}},
  \bibinfo{author}{\bibfnamefont{G.}~\bibnamefont{Liu}},
  \bibinfo{author}{\bibfnamefont{M.}~\bibnamefont{Steger}},
  \bibinfo{author}{\bibfnamefont{D.~W.} \bibnamefont{Snoke}},
  \bibinfo{author}{\bibfnamefont{R.}~\bibnamefont{Balili}},
  \bibinfo{author}{\bibfnamefont{K.}~\bibnamefont{West}}, \bibnamefont{and}
  \bibinfo{author}{\bibfnamefont{L.}~\bibnamefont{Pfeiffer}},
  \bibinfo{journal}{Phys. Rev. X} \textbf{\bibinfo{volume}{3}},
  \bibinfo{pages}{041015} (\bibinfo{year}{2013}).


\bibitem[{\citenamefont{Steger et~al.}(2016)\citenamefont{Steger, Fluegel,
  Alberi, Snoke, Pfeiffer, West, and Mascarenhas}}]{Steger2016}
\bibinfo{author}{\bibfnamefont{M.}~\bibnamefont{Steger}},
  \bibinfo{author}{\bibfnamefont{B.}~\bibnamefont{Fluegel}},
  \bibinfo{author}{\bibfnamefont{K.}~\bibnamefont{Alberi}},
  \bibinfo{author}{\bibfnamefont{D.~W.} \bibnamefont{Snoke}},
  \bibinfo{author}{\bibfnamefont{L.~N.} \bibnamefont{Pfeiffer}},
  \bibnamefont{and} \bibinfo{author}{\bibfnamefont{K.}~\bibnamefont{West}},
 \bibinfo{author}{\bibfnamefont{A.}~\bibnamefont{Mascarenhas}},
  \bibinfo{journal}{Optics Letters}  \textbf{\bibinfo{volume}{42}},
\bibinfo{pages}{1165} (\bibinfo{year}{2017}).

\bibitem[{\citenamefont{Caputo et~al.}(2016)\citenamefont{Caputo, Ballarini,
  Dagvadorj, Sanchez~Munoz, De~Giorgi, Dominici, West, Pfeiffer, Gigli, Laussy
  et~al.}}]{Caputo2016}
\bibinfo{author}{\bibfnamefont{D.}~\bibnamefont{Caputo}},
  \bibinfo{author}{\bibfnamefont{D.}~\bibnamefont{Ballarini}},
  \bibinfo{author}{\bibfnamefont{G.}~\bibnamefont{Dagvadorj}},
  \bibinfo{author}{\bibfnamefont{C.}~\bibnamefont{Sanchez~Munoz}},
  \bibinfo{author}{\bibfnamefont{M.}~\bibnamefont{De~Giorgi}},
  \bibinfo{author}{\bibfnamefont{L.}~\bibnamefont{Dominici}},
  \bibinfo{author}{\bibfnamefont{K.}~\bibnamefont{West}},
  \bibinfo{author}{\bibfnamefont{L.}~\bibnamefont{Pfeiffer}},
  \bibinfo{author}{\bibfnamefont{G.}~\bibnamefont{Gigli}},
  \bibinfo{author}{\bibfnamefont{F.}~\bibnamefont{Laussy}},
  \bibnamefont{et~al.}, \bibinfo{journal}{arXiv:1610.05737}
  (\bibinfo{year}{2016}).

\bibitem[{\citenamefont{Carussoto and Ciuti}(2013)}]{Carusotto13a}
\bibinfo{author}{\bibfnamefont{I.}~\bibnamefont{Carussoto}} \bibnamefont{and}
  \bibinfo{author}{\bibfnamefont{C.}~\bibnamefont{Ciuti}},
  \bibinfo{journal}{Rev. Mod. Phys.} \textbf{\bibinfo{volume}{85}},
  \bibinfo{pages}{299} (\bibinfo{year}{2013}).

\bibitem[{\citenamefont{Doan et~al.}(2005)\citenamefont{Doan, Cao, Thoai, and
  Haug}}]{Doan2005}
\bibinfo{author}{\bibfnamefont{T.~D.} \bibnamefont{Doan}},
  \bibinfo{author}{\bibfnamefont{H.~T.} \bibnamefont{Cao}},
  \bibinfo{author}{\bibfnamefont{D.~B.~Tran} \bibnamefont{Thoai}},
  \bibnamefont{and} \bibinfo{author}{\bibfnamefont{H.}~\bibnamefont{Haug}},
  \bibinfo{journal}{Phys. Rev. B} \textbf{\bibinfo{volume}{\textbf{72}}},
  \bibinfo{pages}{085301} (\bibinfo{year}{2005}).

\bibitem[{\citenamefont{Bobrovska and Matuszewski}(2015)}]{Bobrovska2016}
\bibinfo{author}{\bibfnamefont{N.}~\bibnamefont{Bobrovska}} \bibnamefont{and}
  \bibinfo{author}{\bibfnamefont{M.}~\bibnamefont{Matuszewski}},
  \bibinfo{journal}{Phys. Rev. B} \textbf{\bibinfo{volume}{92}},
  \bibinfo{pages}{035311} (\bibinfo{year}{2015}).

\bibitem[{not({\natexlab{d}})}]{note:note3}
\bibinfo{note}{See details in the Supplemental Material and Fig. S4}.

\bibitem[{\citenamefont{Colas and Laussy}(2016)}]{Colas2016}
\bibinfo{author}{\bibfnamefont{D.}~\bibnamefont{Colas}} \bibnamefont{and}
  \bibinfo{author}{\bibfnamefont{F.~P.} \bibnamefont{Laussy}},
  \bibinfo{journal}{Phys. Rev. Lett.} \textbf{\bibinfo{volume}{116}},
  \bibinfo{pages}{026401} (\bibinfo{year}{2016}).

\end{thebibliography}

\balancecolsandclearpage

\end{document}